"entire.tex paper "
\documentstyle[12pt]{article}

\begin{document}
\baselineskip .25in
\newcommand{\numero}{SHEP 95-37 }   

\newcommand{\titre}{THE CHIRAL 2-SPHERE}
\newcommand{\auteura}{K.J. Barnes}
\newcommand{\auteurb}{J.M. Generowicz}
\newcommand{\auteurc}{P.J. Grimshare}
\newcommand{\place}{Department of Physics,\\
University of Southampton\\
Southampton SO17 1BJ \\ U.K. }
\newcommand{\beq}{\begin{equation}}
\newcommand{\eeq}{\end{equation}}

\newcommand{\abstrait}{The two dimensional surface of a sphere can be
parametrized by
coordinates representing two charged pions acting as Goldstone bosons
of a
broken $SU_2$
symmetry.  We construct in full concrete detail, and in a general
class of
coordinate systems, all
the relevant structure forming a framework for this low energy
effective
theory.}
\begin{titlepage}
\hfill \numero  \\

\vspace{.5in}
\begin{center}
{\large{\bf \titre }}
\\ by   \\ \auteura \\
\auteurb \\
\auteurc \\ \bigskip
\place \bigskip \\

\vspace{.9 in}
{\bf Abstract}
\end{center}
\abstrait
 \bigskip \\
\end{titlepage}

\section{Introduction} It is now some 25 years since non-linear
chiral
$SU_2 \times SU_2$
Lagrangians were introduced to study the experimental consequences of
the
emergence of three
massless pions as Goldstone bosons, and the results have been clearly
exhibited in excellent
review
articles \cite{{1},{2}}.
  Later a very detailed and remarkably successful effective chiral
Lagrangian perturbative
treatment of low energy physics was proposed by Gasser and Leutwyler
\cite{{3},{4}} and is
now regarded as standard in the field.
  In such schemes the transformations of the Goldstone bosons are
non-linear and general
treatments of the required coset-space mathematics are well
established and
elegant in form
\cite{{5},{6}}.
  Also, the consequential construction of invariant non-linear
Lagrangians
is standard and well
known \cite{{7},{8}}.

{}From time to time, as in the case of effective chiral Lagrangians
mentioned
above, there are
developments in physics which create a resurgence of interest in the
structure.
  This was particularly the case when supersymmetric $\sigma$ models
were
first taken seriously
\cite{{9},{10}} because of similarities of their properties in two
dimensions with the structure
of four-dimensional gauge theories \cite{11}.
  The generalization to $CP_N$ models in four dimensions \cite{12}
followed
swiftly, and a
seminal paper by Zumino \cite{13} showed the central place of
geometry in
the models, with the
K\"{a}hler metric of complex manifolds providing an elegant
description of
the supersymmetry.
  There then followed a decade in which the main focus of attention
was on
preon like models in
which the dominant theme was that the supersymmetry helped to ensure
the
existence of light
fermions by relating them to bosons which were in turn kept light by
the
Goldstone theorem.
  A general analysis of the required features can be obtained by
working
backwards through the
literature from the references given in papers by Kotcheff and Shore
\cite{14}, and by
Buchmuller
and Lerche \cite{15}, both of which are written with authority and
also
have fine introductory
sections.

Recently there have been two developments which suggest a yet further
resurgence of interest
in
these topics.
The electric-magnetic duality conjectured by Olive and Montonen
\cite{16}
several years ago,
and
shown by Osborn \cite{17} to be related to $N=4$ supersymmetric gauge
theories, has emerged
in a generalization in the work of Sen \cite{18}.
Moreover, this seems to play an important role in the work of Seiberg
and
Witten \cite{19}
involving $N=2$ supersymmetric gauge theory in four dimensions.
On an apparently unrelated front, following the emergence of the
Supersymmetric Standard
Model as a major candidate for physics beyond the Standard Model, has
come
the realization that
supersymmetry may appear in nature at energies which may soon be
experimentally accessible.
Thus a supersymmetric extension of chiral perturbation theory becomes
of
real interest.
Already, two attempts have been made in this direction
\cite{{20},{21}}
both based on linear
supersymmetric models in which the symmetry is broken (but the
supersymmetry preserved) as
the Higgs mass becomes infinite.

It seems that supersymmetric sigma models are ripe for further
investigation, and obviously
the simplest underlying K\"{a}hler manifold is the 2-sphere
\cite{22}.
What is presented in this paper is a direct treatment of the manifold
structure, the nonlinear
transformation laws of the Goldstone bosons, and the construction of
the
invariant Lagrangians,
all in a general class of coordinate systems.
Curiously, although the 2-sphere has been much studied this does not
seem
to have been recorded
before.
There are, of course, versions in coordinates resulting from
constrained
linear $\sigma$ models,
treatments in exponential (standard) coordinates, projective
coordinate
presentations, and most
importantly stereographic coordinate representations revealing the
K\"{a}hler structure.
Our general coordinate treatment includes and relates all of these,
and we
believe it reveals the
structure in much the way that covariant notation clarifies special
relativity.

We shall show how this model, although not physical, is uniquely
embedded
in chiral $SU_2
\times SU_2$ (which is indeed of direct physical interest as noted
above)
and retains many of the
relevant features thus allowing them to be studied in a much simpler
and
concrete way.
It is a very useful theoretical laboratory.

One of the primary motivations for the current presentation, it has
to be
admitted, is the
experience of one of the authors (K.J.B.) who over many years has
persisted
in quoting some of
the results as ``self-evident'' consequences of the embedding in
chiral
$SU_2 \times SU_2$
which
is fully analysed \cite{23}.  Only repeated objections of friends and
colleagues have finally
persuaded him that ``obvious'' is not equivalent to mathematically
proved.
This paper shows that
even though the techniques used in reference \cite{23} are not all
valid in
the present case,
nevertheless the results obtained in the embedded limit are
identical.

\section{The Chiral Sphere}

We start this section by reviewing \cite{23} the structure of chiral
$SU_2
\times SU_2$ to
establish notation. The transformation of the fundamental (quark)
multiplet
is specified by
\beq
q \rightarrow q - i \theta_i \frac{\tau^i}{2} q - i \phi_i
\frac{\tau^i}{2}
( i \gamma_5)q
\label{1}
\eeq
to lowest order in the real parameters $\theta_i$ and $\phi_i$, $1
\leq i
\leq 3$, where $\tau^i$
are the familiar Pauli matrices.
Note the extra $(i \gamma_5$) factors in the final terms which are
included
to ensure that the
Goldstone bosons of this scheme will be pseudoscalar.
The crucial step in describing these bosons is to parametrize the
coset
space defined by the
quotient of the $SU_2 \times SU_2$ by the vector $SU_2$ parameterised
by
the $\theta^i$ alone.
This takes the simple form
\beq
\hat{L} = \exp \left\{ \frac{-i \theta}{2} n_i \tau^i ( i \gamma_5)
\right\} \label{2}
\eeq
where the Goldstone fields are described by
\beq
M^i = M n^i , \label{3}
\eeq
with
\beq
(n^i)^2 = 1 , \label{4}
\eeq
so that
\beq
(M^i)^2 = M^2 , \label{5}
\eeq
and $\theta$ is an arbitrary function of $M$.
This arbitrariness may be viewed as the freedom to change coordinate
systems on the coset
space,
or to redefine the field variables describing the mesons.
If we define projection operators by
\beq
P_L = {\scriptstyle \frac{1}{2}} ( 1 + i \gamma_5) , \label{6}
\eeq
and
\beq
P_R =  {\scriptstyle \frac{1}{2}} ( 1 - i \gamma_5) , \label{7}
\eeq
so that
\beq
P_L P_L = P_L , \label{8}
\eeq
\beq
P_R P_R = P_R , \label{9}
\eeq
\beq
P_L P_R = 0 = P_R P_L , \label{10}
\eeq
and
\beq
P_L + P_R = 1 , \label{11}
\eeq
then we can rewrite equation (\ref{2}) as
\beq
\hat{L} = L P_L + L^{-1} P_R , \label{12}
\eeq
where $L$ is unitary and the $\gamma_5$ dependence is now contained
solely
in the projection
operators.
It is then clear that we can deal with
\beq
L = \exp \left\{ \frac{ - i \theta}{2} n_i \tau^i \right\}
\label{13}
\eeq
and reinstate the $\gamma_5$ factors only when wishing to consider
the
explicit couplings of
the
Goldstone bosons to matter fields.
The action of a group element $g$ (of $SU_2 \times SU_2$) on the
coset
space can be specified
by
\beq
gL = L ' h \label{14}
\eeq
where
\beq
L' (M_i ) = L ( {M '}_i ) , \label{15}
\eeq
specifies the non-linear transformations of the Goldstone boson
fields,
\beq
h = \exp \left\{ \frac{ - i}{2} \lambda_i \tau^i \right\} ,
\label{16}
\eeq
and the $\lambda_i$ depend on the fields and the group parameters.
What we have are non-linear transformations among the $M_i$ (which
give a
realization of the
group) which are linear under the action of the $SU_2$ subgroup, thus
neatly describing a
situation where the full group is still realized, but  in a manner
well
suited to spontaneous breaking
to the subgroup.
The Goldstone bosons are a linear representation of the $SU_2$
subgroup
only.
Although the procedure extends to other representations, for our
present
purposes it will be
sufficient to stay mostly in the fundamental representation.

We are now ready to discuss the chiral $SU_2$ structure embedded in
this
framework.  Consider
the subgroup of the chiral $SU_2 \times SU_2$ group specified in
equation
(\ref{1}) by retaining
only the parameters $\theta_3$ and $\phi_A$, with $A = 1$ and $2$.
Obviously this is an $SU_2$ subgroup, and we call it chiral $SU_2$ in
recognition of the
$(i\gamma_5)$ factors with the $\tau^A$ generators.
Clearly the $\tau^3$ generates a $U_1$ subgroup, so that the coset
space
obtained by the
quotient of chiral $SU_2$ by this $U_1$ is parametrized by
coordinates $M_A
$, $A = 1$ and
$2$, which
can be viewed as describing two Goldstone pseudoscalars.
Notice that the embedding of this $SU_2/U_1$ structure in the
$\frac{SU_2
\times
SU_2}{SU_2}$ structure is uniquely specified.
Moreover, if we set $M_3$ and $n_3$ to zero in our previous
discussion,
then
\beq
L = \exp \left\{ \frac{- i \theta}{2} n_A \tau^A \right\} ,
\label{17}
\eeq
and set $\lambda_A = 0$, so
\beq
h = \exp \left\{ \frac{ - i}{2} \lambda_3 \tau_3 \right\} ,
\label{18}
\eeq
where $\theta$ is now an arbitrary function of
\beq
M^2 = M_1^2 + M_2^2 \label{19}
\eeq
which when $M_3$ becomes zero remains as the only independent scalar.

We can now see the advantages of using this chiral 2-sphere as a
model.  It
is simpler than the
chiral $SU_2 \times SU_2$ scheme even in the purely bosonic sector.
Moreover, the 2-sphere is a K\"{a}hler manifold and so admits a
supersymmetric extension in
which
the Goldstone bosons acquire fermionic (Weyl) partners without yet
more
quasi-Goldstone
bosons and fermions being forced into the model \cite{22}.
Also the resulting couplings among the particles are uniquely
specified.
Contrast this with the situations in references \cite{20} and
\cite{21}
where the number of
bosons
doubles, as does the number of associated fermions, and finally the
couplings involving these
new
particles are not uniquely specified.
Of course, these latter cases are closer to the physics of the real
world
(they have 3 pions for
example), but the embedded chiral 2-sphere model retains many
significant
features and is a far
more tractable theoretical laboratory.
We now present the details of this model.

First we establish the transformation laws of the Goldstone fields
under
chiral $SU_2$.
It is sufficient to work to lowest order in the group parameters and
we
denote the transformations
by
\beq
g: M_A \rightarrow M_A + \theta_3 K_{3A} + \phi_B K_{BA} , \label{20}
\eeq
where $K_{3A}$ and $K_{BA}$ are Killing field components constructed
from
the $M_A$
themselves.
Of course, the action under an element of the $U_1$ subgroup is
linear so
that $K_{3A}$ is
already known, but we shall let this emerge from our calculations.
Expanding equation (\ref{14}) we see that we need to solve
\beq
\begin{array}{l}
\left[ 1 - \frac{i \theta_3}{2} \tau_3 - \frac{i\phi_B}{2} \tau_B
\right]
L(M) \\ \\
= \left[ L(M) + {L,}_A \theta_3 K_{3A} + {L,}_A \phi_B K_{BA} \right]
\times
\left[ 1 - \frac{i}{2} \theta_3 \tau_3 - \frac{i \phi_A
\lambda_{A3}}{2}
\tau_3 \right] \; ,
\end{array} \label{21}
\eeq
where
\beq
{L,}_A = \frac{ \partial L(M)}{\partial M_A} , \label{22}
\eeq
\beq
\lambda_3 = \theta_3 + \phi_A \lambda_{A3} , \label{23}
\eeq
and we note that in this particular simple example raising and
lowering of
indices is of no
consequence if we preserve the order of indices on the Killing vector
fields.
It is clear that the calculations require nothing more than the
construction of functions of Pauli
matrices, but even so a little technique can be helpful.
The quantities
\beq
P^{\pm} = {\scriptstyle \frac{1}{2}} ( 1 \pm n_A \tau_A ) \label{24}
\eeq
share the projection operator properties given in equations (\ref{8})
to
(\ref{11}) for the $P_L$
and $P_R$, as can easily be seen because the $n_A$ form a unit
vector.
This means that equation (\ref{17}) can be expressed as
\beq
L = P^+ \exp \left(  \frac{-i \theta}{2} \right) + P^- \exp \left(
\frac{i\theta}{2} \right) ,
\label{25}
\eeq
and other functions can be similarly handled.
Also, from equation (\ref{19}) we see that
\beq
{M ,}_A = n_A , \label{26}
\eeq
and differentiating
\beq
M_A = M n_A \label{27}
\eeq
yields
\beq
M n_{A,B} = \delta_{AB} - n_A n_B , \label{28}
\eeq
so that
\begin{eqnarray}
{P,}_B^{\pm} &=& \pm {\scriptstyle \frac{1}{2 M}} \tau_A (\delta_{AB}
- n_A
n_B)
\nonumber \\
&=& \pm {\scriptstyle \frac{1}{2M}} \left( \tau_B + n_B P^- - n_B P^+
\right) . \label{29}
\end{eqnarray}
We note that again the tensors $(\delta_{AB} - n_A n_B)$ and $n_A
n_B$ have
the by now
familiar projection operator properties, so that calculations become
systematic and
straightforward.
A little simple algebra applied to equation (\ref{21}) reveals that
\beq
K_{BC} = M \cot \theta ( \delta_{BC} - n_B n_C) + n_B n_C \frac{dM}{d
\theta}
\label{30}
\eeq
and
\beq
K_{3C} = \varepsilon_{3BC} n_B \phi = \varepsilon_{3BC} M_B ,
\label{31}
\eeq
where $\varepsilon_{3BC}$ is the familiar totally antisymmetric
Levi-Civita
tensor.  As noted
previously $K_{3C}$ is linear in the $M_C$, and we recognise the
usual
rotational
transformation
of a vector.

We have already found the transformation laws for the Goldstone
bosons and,
as the reader can
easily check, these are identical to those given in reference
\cite{23}
when the truncation of
variables described above is applied.  Returning to equations
(\ref{14})
and (\ref{18}) we note,
following reference \cite{5}, that if $\psi$ is an irreducible
representation of the unbroken
subgroup, so that here (keeping to the fundamental representation) we
have
simply that
\beq
\psi \rightarrow \psi - \frac{i \theta_3}{2} \tau_3 \psi \label{32}
\eeq
then under the full group action
\beq
\psi \rightarrow \psi - \frac{i \theta_3}{2} \tau_3 \psi - \frac{i
\phi_B
\lambda_{B3}}{2}
\tau_3 \psi \label{33}
\eeq
where
\beq
\lambda_{B3} = \varepsilon_{BA3} M_A \tan (\theta / 2) /M .
\label{34}
\eeq
Note that this transformation law is linear in $\psi$, but with
non-linear
coefficients constructed
from $M_A$; it and its generalizations are known as Standard Field
transformations, and these
exhaust all field types.
Again the reader can easily check that the result in equation
(\ref{34})
follows trivially from the
corresponding result in reference \cite{23} when our truncation
method is
applied.

What remains is to show how to construct invariant Lagrangians from
the
fields we have
introduced.
It is at this point that the objections (mentioned previously in the
Introduction) arise to the direct
extraction of further results from reference
\cite{23} by our truncation method.
The difficulty is that later results in reference \cite{23}
explicitly use
a property that is not
available in the chiral $SU_2$ substructure.
In the full chiral $SU_2 \times SU_2$ the Killing vectors can be
combined
into so called left and
right combinations which viewed as matrices $\left( K^L \right)_{AB}$
and
$\left( K^R
\right)_{AB}$ are non-singular and can be inverted.
Unfortunately, in the chiral $SU_2$ substructure only $K_{AB}$ and
$K_{3C}$
exist so that
this
trick (which is a useful shortcut) is not directly available.
However, as we shall see, $K_{AB}$ itself is non-singular, and by a
slight
extension of the
calculations we do eventually reach the same results.

So what invariants can be constructed?  This question was answered
elegantly in reference
\cite{7}.
The first point is that no invariant can be constructed from the
$M_A$
alone.
In particular this implies that an invariant mass term is not
available for
the Goldstone bosons in
accordance with the Goldstone theorem.
Now consider derivatives of the fields.
The key concept is found by rewriting equation (\ref{14}) in the
forms
\beq
L' = gL h^{-1}, \label{35}
\eeq
and
\beq
{L'}^{-1} = h L^{-1} g^{-1} , \label{36}
\eeq
and differentiating the former to obtain
\beq
\partial_\mu L' = g \left[ ( \partial_\mu L) h^{-1} + L (
\partial_\mu
h^{-1} ) \right] , \label{37}
\eeq
where $\partial_\mu = \frac{\partial}{\partial x^\mu}$ differentiates
the
fields $M_A$, but $g$
is constant because we are considering only global transformations.
{}From equations (\ref{36}) and (\ref{37}) we see
\begin{eqnarray}
L^{-1} \left( \partial_\mu L \right) & \rightarrow & {L '}^{-1}
\left(
\partial_\mu L ' \right)
\nonumber \\
 & = & h \left[ L^{-1}  \left( \partial_\mu L \right) \right] h^{-1}
+ h
\left( \partial_\mu h^{-1}
\right) \label{38}
\end{eqnarray}
and recognise that, because $h$ is in the subgroup, the
transformation does
not mix the coset
space and subgroup generators in the algebra.
Thus, if we write
\begin{eqnarray}
2i L^{-1} \left( \partial_\mu L \right) & = & \tau_B a_\mu^B + \tau_3
v_\mu^3 \nonumber \\
 & = & a_\mu + v_\mu \label{39}
\end{eqnarray}
then equation (\ref{38}) gives
\beq
a_\mu \rightarrow h a_\mu h^{-1} \label{40}
\eeq
and
\begin{eqnarray}
v_\mu & \rightarrow & h v_\mu h^{-1} + h \left( \partial_\mu h^{-1}
\right)
\nonumber \\
 & = & v_\mu + h \left( \partial_\mu h^{-1} \right) \label{41}
\end{eqnarray}
where the final simplification in equation (\ref{41}) follows because
the
subgroup is abelian.
It follows from equation (\ref{40}) that the quantity
\[
\frac{1}{2} Tr \left[ a_\mu^B {a^\mu}_B \right]
\]
 is an invariant, and in fact this is the only invariant which can be
made
from the Goldstone
bosons
which involves exactly two derivatives.
Usually the notation of a covariant derivative
\beq
\Delta_\mu M^B = a_\mu^b \label{42}
\eeq
is introduced and the expression
\beq
{\cal L} =  {\scriptstyle \frac{1}{2}} Tr [ ( \Delta_\mu M^B) (
\Delta^\mu
M_B) ]
\label{43}
\eeq
written for the Lagrangian which has a leading order expansion in
fields
appropriate for
interpretation as a kinetic energy term.
Isham \cite{8} introduced the metric form
\beq
{\cal L} = {\scriptstyle \frac{1}{2}} g_{AB} ( \partial_\mu M^A ) (
\partial^\mu M^B )
\label{44}
\eeq
for this Lagrangian, thus giving a geometric understanding in terms
of the
metric $g_{AB}$ on
the coset space manifold.
We return briefly to equation (\ref{41}) to note that if there is a
matter
field $\psi$ which
transforms under the $U_1$ subgroup so that
\beq
\psi \rightarrow \psi - \frac{i}{2} \theta_3 \tau^3 \psi \label{45}
\eeq
then reference \cite{5} shows that under the full action of the
chiral
$SU_2$
\beq
\psi \rightarrow \psi -  \frac{i \theta_3}{2} \tau_3 \psi - \frac{i
\lambda_{A3} \phi_A}{2} \tau_3
\psi \label{46}
\eeq
and so
\beq
\Delta_\mu \psi = \partial_\mu \psi - \frac{i}{2} v_\mu^3 \tau^3 \psi
\label{47}
\eeq
is a covariant derivative transforming as $\psi$ itself in equation
(\ref{46}), and may be used to
form invariant terms involving matter fields in the usual way
\cite{{5},{7}}.

In the remainder of this paper we derive expressions for the
covariant
derivatives and metric by
direct manipulation of the Pauli matrices, and remaining strictly
within
the chiral $SU_2$
framework.
We start by introducing a little extra calculational device by
defining
\beq
R_{ij} = \frac{1}{2} Tr [ L^{-1} \tau_i L \tau_j ]
\label{48}
\eeq
where, as before, $i$ and $j$ lie in the range $1-3$.  Using the same
formalism as in equations
(\ref{21}) to (\ref{29}), we easily establish that
\begin{eqnarray}
R_{AB} &=& ( \delta_{AB} - n_A n_B ) \cos \theta + n_A n_B ,
\label{49} \\
R_{A3} &=& \varepsilon_{AB3} n_B \sin \theta = - R_{3A} , \label{50}
\end{eqnarray}
and
\beq
R_{33} = \cos \theta , \label{51}
\eeq
where the projection operator properties are again noted.  From
equation
(\ref{35}) we see that
the quantities appearing in the covariant derivatives can be
expressed as
\beq
a_{\mu B} = ( \partial_\mu M_C) a_{CB} \label{52}
\eeq
and
\beq
v_{\mu 3} = ( \partial_\mu M_C ) v_{C3} \label{53}
\eeq
where
\beq
a_{CB} = i Tr [ \tau_B L^{-1} {L,}_C ] \label{54}
\eeq
and
\beq
v_{C3} = i Tr [ \tau_3 L^{-1} {L,}_C ] \label{55}
\eeq
which we shall shortly see are particularly convenient forms.
Now we return to our defining equation (\ref{21}) and extract
\beq
\frac{-i}{2} \tau_A L = {L,}_B K_{AB} - \frac{i}{2} \lambda_{A3} L
\tau_3 ,
\label{56}
\eeq
and we can deduce that
\beq
R_{AD} = K_{AB} a_{BD} \label{57}
\eeq
by premultiplying by $\tau_D L^{-1}$ and taking the trace.
Since $K_{AB}$ is non-singular, we can see that
\beq
a_{FD} = \left( K^{-1} \right)_{FA} R_{AD} , \label{58}
\eeq
and hence
\beq
a_{FD} = ( \delta_{FD} - n_F n_D ) \frac{\sin \theta}{M} + n_F n_D
\frac{d
\theta}{d M}
\label{59}
\eeq
follows from equations (\ref{30}) and (\ref{49}).
Similarly, returning to equation (\ref{56}) we can also deduce that
\beq
R_{A3} = K_{AB} v_{B3} + \lambda_{A3}
\label{60}
\eeq
by premultiplying by $\tau^3 L^{-1}$ and tracing.
Hence we find directly that
\beq
v_{F3} = \frac{2}{M} \sin^2 (\theta/2) \varepsilon_{F Z 3} n_Z
\label{61}
\eeq
by using equations (\ref{34}) and (\ref{50}).
This completes our task, and we see that all the results can indeed
be
found from those in
reference (\ref{23}) by our truncation method.
We do realize that we have not given a strict mathematical proof of
the
relationship between
chiral
$\frac{SU_2 \times SU_2}{SU_2}$ and the chiral $SU_2 / U_1$ embedded
in it.

It is however gratifying to see that all the results we need do come
out as
speculated in the
truncation.

One last footnote.
Just as in general relativity where tetrads or vierbeine are
introduced to
allow the treatment of
spinors by ``taking the square root of the metric'', here the unitary
unimodular square root nature
of $L$ versus $L^2$ can be exploited by introducing Killing vectors
for the
square root system.
This concept is easier to understand in concrete form.
{}From our defining equation (\ref{14}) we can see that
\beq
L \stackrel{\sim}{g}^{-1} = h^{-1} L ' \label{62}
\eeq
where we have inverted the equation and then applied the involutive
outer
automorphism $\sim$
which reverses the signs of the generators in the group but not in
the
subgroup.
Multiplying the respective sides of equations (\ref{14}) and
(\ref{62})
gives
\beq
g L^2 \stackrel{\sim}{g}^{-1} = {L '}^2 \label{63}
\eeq
in which $h$ has been eliminated thus emphasizing that the action on
$M_A$,
specified by
$K_{BA}$, is determined by $L^2$.
In the notation used previously we have
\beq
\left\{ \tau_{A,} L^2 \right\} = - 2 L_{, B}^2 K_{AB} \label{64}
\eeq
as the significant part of the information.
We multiply from the left by $L^{-2} \left( K^{-1} \right)_{CA}$ to
see
that
\beq
\left( K^{-1} \right)_{CA}
\left[ L^{-2} \tau_A L^2 + \tau_A \right]
= - 2 i L^{-2} L_{, C}^2 , \label{65}
\eeq
then multiplying from the right by $\frac{1}{2} \tau^B$ and tracing
yields
\beq
\left( K^{-1} \right)_{CA}
\left[ \delta_{AB} + {\scriptstyle \frac{1}{2}} Tr \left( L^{-2}
\tau_A L^2
\tau_B \right) \right]
= - i Tr \left( L^{-2} L_{, C}^2 \tau_B \right) \label{66}
\eeq
and comparison with equations (\ref{48}) and (\ref{54}) makes clear
how the
square root can
be
taken.
We define
\beq
\left( k^{-1} \right)_{CA}
\left[ \delta_{AB} + R_{AB} \right]
=  i Tr \left( \tau_B L^{-1} {L,}_C \right) \label{67}
\eeq
where the sign in taking the square root has been picked for
convenience.
Then equations (\ref{56}) and (\ref{57}) reveal that
\beq
\left( k^{-1} \right)_{QT} =  \left(K^{-1} \right)_{QA}
R_{AD}  \left( \left[ 1 + R \right]^{-1} \right)_{DT} \label{68}
\eeq
which the reader may enjoy confirming, reproduces the obvious inverse
of
$K_{QT}$ in equation
(\ref{30})
when $\theta$ is halved.
This clarifies the sense of the square root.
In an entirely analogous way we may write
\beq
\left(
k^{-1} \right)_{CB}  R_{B3}=  i Tr \left( \tau_3 L^{-1} L'_C \right)
\label{69}
\eeq
and discover
\beq
\lambda_{A3} = R_{A3} - K_{AB} \left( k^{-1} \right)_{BF} R_{F3} .
\label{70}
\eeq
Substitution of the results from equation (\ref{68}) and (\ref{50})
into
equation (\ref{69})
confirms the expression found in equation (\ref{61}) for $v_{F3}$,
while
similar substitutions
into equation (\ref{70}) retrieve the result previously given in
equation
(\ref{34}).
The results given in this last section are not directly retrievable
(as far
as we know) by truncation
of the results in reference \cite{23}, since the full chiral
structure
allowed shortcuts to be taken
in that paper.

\section{Acknowledgements}

This work has been supported in part by SERC (now PPARC) grant number
GR/J21569.

\end{document}